\documentclass[pre,showpacs,showkeys,preprintnumbers,amsmath,amssymb,superscriptaddress,twocolumn,nofootinbib]{revtex4}
\usepackage{graphicx}
\usepackage{amsfonts}
\usepackage{url}
\usepackage{epstopdf}
\usepackage{color}
\usepackage{bm}
\usepackage[colorlinks=true,linkcolor=blue,citecolor=red]{hyperref}%

\def\asin{\mathrm{asin}}

\begin{document}

\title{Mean first-passage time to a small absorbing target in an elongated planar domain}
\author{Denis~S.~Grebenkov}
 \email{denis.grebenkov@polytechnique.edu}
\affiliation{
Laboratoire de Physique de la Mati\`{e}re Condens\'{e}e (UMR 7643), \\ 
CNRS -- Ecole Polytechnique, IP Paris, 91128 Palaiseau, France}

\author{Alexei T. Skvortsov}
\affiliation{Maritime Division, Defence Science and Technology, 506 Lorimer Street, Fishermans Bend, Victoria 3207, Australia}

\date{\today}

\begin{abstract}
We derive an approximate but fully explicit formula for the mean
first-passage time (MFPT) to a small absorbing target of arbitrary
shape in a general elongated domain in the plane.  Our approximation
combines conformal mapping, boundary homogenisation, and Fick-Jacobs
equation to express the MFPT in terms of diffusivity and geometric
parameters.  A systematic comparison with a numerical solution of the
original problem validates its accuracy when the starting point is not
too close to the target.  This is a practical tool for a rapid
estimation of the MFPT for various applications in chemical physics
and biology.
\end{abstract}





\maketitle

\section{Introduction}

The concept of first-passage time (FPT) is ubiquitous in describing
phenomena around us.  Being originally stemmed from the theory of
Brownian motion as a time taken for a diffusing particle to arrive at
a given location, nowadays it is widely used in chemistry
(geometry-controlled kinetics), biology (gene transcription, foraging
behaviour of animals) and many applications (financial modelling,
forecasting of extreme events in the environment, time to failure of
complex devices and machinery, military operations).  This subject has
an extensive literature, see
\cite{Redner_2001,Bray_2013,Metzler_2014,Benichou_2014,Bressloff_2015,Holcman_2015,Lindenberg_2019,Dagdug_2015,Oshanin_2009,Lindsay_2017,Grebenkov_2016,Skvortsov_2018,Skvortsov_2020,Grebenkov_2020,Grebenkov_2019} and references therein. 
There is also a rich variety of different physical phenomena that can
be analytically treated with the framework of the FPT due to
similarity of underlying equations \cite{Bazant_2016}.

In a basic setting, the first-passage problem is formulated in the
following way.  We consider a Brownian particle initially located at
point $\textbf{r}$ of a bounded domain $\Omega$ and searching for a
small target $\mathcal{S}$ (a small region with absorbing boundary)
inside that domain (if the target is at the boundary the problem is
usually referred to as the narrow escape problem \cite{Holcman_2015}).
As the first-passage time of the particle to the target is a random
variable, its full characterisation requires the computation of its
distribution
\cite{Mejia12,Rupprecht15,Godec_2016a,Godec_2016b,Lanoiselee_2018,Grebenkov_2018a,Grebenkov_2018b,Grebenkov_2019b}.
In many practical situations, however, it is enough to estimate the
average time $T(\textbf{r})$ taken for the particle to hit the target
(see
\cite{Redner_2001,Metzler_2014,Benichou_2014,Condamin_2007,Benichou_2008,Benichou_2010,Guerin_2016}
and references therein).  The mean first-passage time (MFPT) satisfies
the Poisson equation \cite{Redner_2001}
\begin{equation}
 \label{I:e1}
D \Delta T (\textbf{r}) = -1,
\end{equation}
where $D$ is the particle diffusivity, and $\Delta$ is the Laplace
operator.  The boundary of the domain is assumed reflecting,
${\partial T }/{\partial n} =0$ on $\textbf{r} \in \partial \Omega$
(with $\partial/\partial n$ being the normal derivative) and the
target surface is absorbing, $T = 0$ on $\textbf{r} \in \partial
\mathcal{S}$.

In spite of an apparent simplicity of Eq. (\ref{I:e1}) and a variety
of powerful methods for its analysis, to date the exact closed-form
solutions of Eq. (\ref{I:e1}) are available only for a few special
cases and for the domains with high symmetry such as a sphere or a
disk \cite{Redner_2001}.  Many approximate solutions, derived by
advanced asymptotic methods, can produce a remarkable agreement with
numerical solutions of Eq.~(\ref{I:e1}), but often require specific
mathematical expertise and still involve some level of numerical
treatment
\cite{Metzler_2014,Benichou_2014,Holcman_2015,Lindsay_2017,Grebenkov_2016,Singer06a,Singer06b,Singer06c,Pillay10,Cheviakov10,Cheviakov12,Caginalp12,Marshall16,Ward_2020}.
This necessitates the development of analytical approximations that
being perhaps less accurate can lead to simple explicit expressions
that provide reasonable estimations for MFPT in some general geometric
settings.  This was one of the main motivations for the present study.

The aim of the paper is to derive a general formula for the MFPT in an
{\it elongated} planar domain with reflecting boundaries.  The profile
of the domain is assumed to be smooth, slowly changing, but otherwise
general.  The target is assumed to be small but of an arbitrary shape.
We validate our findings by numerical solution of Eq. (\ref{I:e1}) via
a finite elements method.  Remarkably, this simple general formula,
derived under a number of simplified approximations, turns out to be
surprisingly accurate.

\begin{figure}
\centering
\includegraphics[height=0.30\textwidth]{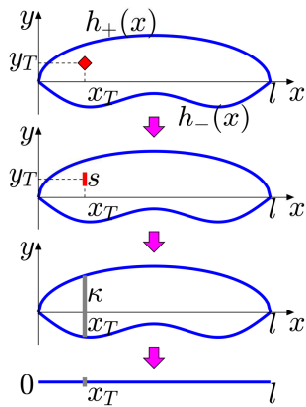}
\caption{
Three-step transformation of the original MFPT problem into an
effectively one-dimensional problem with semi-permeable semi-absorbing
boundary condition.}
\label{fig:scheme}
\end{figure}

\section{Approximation for the MFPT} 

We consider an elongated planar domain of ``length'' $l$, which is
determined by two smooth profiles $h_-(x) < h_+(x)$
(Fig. \ref{fig:scheme}):
\begin{equation}
\Omega = \{(x,y)\in {\mathbb R}^2 ~:~ 0<x<l, ~ h_-(x) < y < h_+(x)\}.
\end{equation}
In particular, the local ``width'' of the domain is $h(x) = h_+(x) -
h_-(x)$, and $h_0 = \max\{ h(x)\}$ is the maximal width.  Throughout
the paper, we assume that the aspect ratio $h_0/l$ of the domain is
small.  A small absorbing target (also called trap or sink) is located
inside the domain at some $(x_T,y_T)$, which is not too close to the
boundary.

The main analytical formula will be derived by employing a three-step
approximation.  First, we replace the absorbing target by a vertical
absorbing interval of the same trapping rate or the same conformal
radius or the same logarithmic capacity (since they all are
proportional to each other \cite{Landkof_1978}).  Far away from the
target such a replacement is justifiable since at the distance greater
than the size of the target (but still much smaller than $h(x_T)$ and
$l$) the absorption flux can be characterised by the first (monopole)
moment of the shape of the target, and this equivalence preserves it.
For a variety of planar shapes (circle, ellipse, arc, triangle,
square, or even some fractals \cite{Ransford_2007,Ransford_2010}),
conformal radius is well-known or can be accurately estimated from
various approximations, see
\cite{Skvortsov_2018,Landkof_1978,Ransford_2010} and references
therein.  For instance, for an elliptical target with semi-axes $a$
and $b$, its conformal radius is $(a+b)/2$, so for an absorbing
interval of length $s$ conformal radius is simply $s/4$ and this
allows us to deduce the length of the equivalent absorbing interval
$s$ for any given target.

Second, we substitute the absorbing interval at $x = x_T$ by an
equivalent semi-permeable semi-absorbing vertical boundary.  In line
with the conventional arguments of effective medium theory, the
trapping effect of the target can approximately be captured by means
of this boundary with an effective reactivity $\kappa$.  More
specifically, we assume that the trapping rate of this effective
boundary is equal to the trapping flux of the particles induced by the
presence of the target.  The well-known examples of such an approach
are acoustic impedance of perforated screens
\cite{Crocker_1998} and effective electric conductance of lattices and
grids \cite{Tretyakov_2003,Hewett_2016,Marigo_2016}.  The effective
reactivity can be related to the geometrical parameters by employing
the ideas of boundary homogenisation
\cite{Skvortsov_2020,Hewett_2016,Marigo_2016}.  In particular, an
explicit form of this reactivity in the case of two absorbing arcs on
the reflecting boundary of a disk of radius $R$ was found in
\cite{Skvortsov_2020}.  As shown in \cite{Skvortsov_2020},
an appropriate conformal mapping allows one to transform such a disk
into an infinite horizontal stripe of width $2h$ with reflecting
boundary that includes two identical absorbing intervals.  By
symmetry, this domain is also equivalent to a twice narrower stripe
(i.e., of width $h$) with a single absorbing interval with a
prescribed offset with respect to the reflecting boundary.  Upon these
transformations, the original formula of the effective reactivity is
preserved, except that the perimeter of the disk, $2\pi R$, is
replaced by the stripe width:
\begin{equation}
\label{M:e9}
 \kappa = \frac{D}{h(x_T)} \frac{\pi}{\ln {(1/F)}} \,,
\end{equation}
where
\begin{equation}
\label{M:e10}
F = \sqrt{\sin^2 [\tfrac{\pi}{2} (\sigma + \sigma_g)] - \sin^2 (\tfrac{\pi}{2} \sigma_g)  } \,,
\end{equation}
with $\sigma = s/h(x_T)$ and $\sigma_g = (y_T - s/2)/h(x_T)$.  Here,
we used the width of the domain, $h(x_T)$, at the location of the
target.  Even though Eq. (\ref{M:e9}) was derived for an infinite
stripe, it is also applicable for an elongated rectangle of width
$h(x_T)$.  Moreover, we will use it as a first approximation for
general elongated domains.

Third, the Brownian particle, which is released at some point $(x,y)$
inside an elongated domain, frequently bounces from the horizontal
reflecting walls while gradually diffusing along the domain towards
the target.  The shape of the horizontal walls (defined by
$h_{\pm}(x)$) can additionally create an entropic drift, which can
either speed up or slow down the arrival to the target.  In any case,
the information about the particle initial location in the vertical
direction, $y$, becomes rapidly irrelevant, and the original MFPT
problem, Eq. (\ref{I:e1}), is reduced to a one-dimensional problem.
While the classical Fick-Jacobs equation determines the concentration
$c(x,t)$ averaged over the cross-section (see
\cite{Zwanzig_1992,Reguera_2001,Kalinay_2006,Bradley_2009,Rubi_2010,Dagdug_2012}
and references therein), the survival probability is determined by the
backward diffusion equation with the adjoint diffusion operator
\cite{Gardiner}.  In particular, Eq.(\ref{I:e1}) in an elongated
domain reduces to
\begin{equation}  \label{eq:T_FJ}
\frac{D}{h(x)} \frac{d}{dx} \biggl[h(x) \frac{dT (x)}{dx} \biggr] = -1\,.
\end{equation}
In summary, we transformed the original problem of finding the MFPT to
a small target of arbitrary shape in a general elongated domain to the
one-dimensional problem, which can be solved analytically.

\begin{table*}[t]
\begin{center}
\begin{tabular}{|c|c|c|c|c|c|c|}  \hline
Domain & $y(z)$ & $Y(z)$ & $U_-(z)$ & $U_+(z)$ &
$Y(1)$ & $C_y$ \\ \hline
Rectangular & $1$ & $z$ & $\tfrac{1}{2} z^2$ & $U_-(1-z)$ & $1$ & $\tfrac{1}{6}$ \\
Triangular & $z$ & $\tfrac{1}{2} z^2$ & $\tfrac{1}{4} z^2$ & $\tfrac{1}{4}(z^2 - 1 - 2\ln z)$ & $\tfrac{1}{2}$ & $\tfrac{1}{8}$ \\
Rhombic & $1 - |2z -1|$ & $\left\{\begin{array}{ll} z^2 & (z<\tfrac{1}{2}) \\ \tfrac{1}{2} -(1-z)^2 & (z>\tfrac{1}{2}) \\ \end{array} \right.$ & 
$\left\{\begin{array}{ll} \tfrac{1}{4} z^2 & (z<\tfrac{1}{2}) \\ \frac{(1-z)^2 - \ln(2-2z)}{4} & (z>\tfrac{1}{2}) \\ \end{array} \right.$ & 
$U_-(1-z)$  & $\tfrac{1}{2}$ & $\tfrac{3}{32}$ \\
Sinusoidal & $\sin (\pi z)$ & $\frac{1}{\pi} (1-\cos(\pi z))$ & 
$\frac{1}{\pi^2} \bigl(\ln 2 - \ln(1 + \cos(\pi z))\bigr)$ & $U_-(1-z)$ &  $\tfrac{2}{\pi}$ & $\tfrac{1}{\pi^2}$ \\
Parabolic & $4 z ( 1 - z)$ & $2z^2(1-2z/3)$ & $\tfrac{1}{6} \bigl(z^2 - z - \ln(1-z)\bigr)$ & $U_-(1-z)$ 
& $\tfrac{2}{3}$ & $\tfrac{19}{180}$ \\
Elliptic & $\sqrt{1 - (2z - 1)^2}$ & $\begin{array}{l} \tfrac{\pi}{8} + \tfrac{1}{4} \asin(2z-1) \\ - (\tfrac{1}{2}-z)\sqrt{z(1-z)} \\ \end{array}$ & 
$\begin{array}{l} \frac{\pi^2}{64} + \frac{z(z-1)}{4} + \frac{\asin^2(2z-1)}{16} \\  + \frac{\pi}{16} \asin(2z-1) \end{array}$ &
$U_-(1-z)$ & $\tfrac{\pi}{4}$ & $\frac{\pi^2}{48} - \tfrac{5}{64}$ \\   \hline
\end{tabular}
\caption{
Several examples of symmetric elongated domains defined by setting $-
h_-(x) = h_+(x) = \tfrac{1}{2} h(x) = \tfrac{1}{2}\, h_0\, y(x/l)$:
the rescaled profile $y(z)$, its integral $Y(z)$, functions
$U_{\pm}(z)$ from Eq. (\ref{eq:U}), the rescaled area $Y(1)$ (such
that $S = lh_0 Y(1)$), and the shape-dependent constant $C_y$ from
Eq. (\ref{eq:C}).  For all domains, which are symmetric with respect
to the vertical line at $l/2$, one has $U_+(z) = U_-(1-z)$.}
\label{tab:Domains} 
\end{center}
\end{table*}

We search for the solution of Eq. (\ref{eq:T_FJ}) in the intervals
$(0,x_T)$ and $(x_T,l)$.  Multiplying this equation by $h(x)/D$,
integrating over $x$ and imposing Neumann (reflecting) boundary
conditions at $x = 0$ and $x = l$, we get
\begin{equation}  \label{eq:Tx0}
T(x) = \left\{ \begin{array}{l l} \displaystyle C_- - \int\nolimits_0^x dx' \frac{S(x')}{D h(x')} & (0< x < x_T), \\
\displaystyle C_+ -  \int\nolimits_x^l dx' \frac{S(l) - S(x')}{D h(x')} & (x_T < x < l), \\ \end{array} \right.
\end{equation}
where $S(x) = \int\nolimits_0^x dx' \, h(x')$ is the area of
(sub)domain restricted between $0$ and $x$.  The integration constants
$C_\pm$ are determined by imposing the effective semi-permeable
semi-absorbing boundary condition at the target location:
\begin{subequations}
\label{M:e8}
\begin{align}
T(x_T - 0) &= T(x_T + 0), \\  \label{eq:Robin}
D \left [ \frac{d T}{dx}(x_T + 0) - \frac{d T}{dx}(x_T - 0) \right] & = \kappa \,T(x_T) .
\end{align}
\end{subequations}
The first relation ensures the continuity of the MFPT, whereas the
second condition states that the difference of the diffusive fluxes at
two sides of the semi-permeable boundary at $x_T$ is equal to the
reaction flux on the target.  The latter flux is proportional to
$T(x_T)$, with an effective reactivity $\kappa$ from Eq. (\ref{M:e9}).

Substituting Eq. (\ref{eq:Tx0}) into Eqs. (\ref{M:e8}), we get the
final solution of the problem: 
\begin{equation}  \label{eq:Tx_main}
T(x) = \frac{l^2}{D} \biggl(U_{\sigma_x}(x_T/l) - U_{\sigma_x}(x/l)\biggr) + \frac{l}{\kappa} \, \frac{Y(1)}{y(x_T/l)} \,,
\end{equation} 
where $y(z)$ is the rescaled profile, $h(x) = h_0 y(x/l)$ (with $z =
x/l$), and
\begin{align}
Y(z) & = \int\limits_0^z dz' \, y(z'), \\  \label{eq:U}
U_{-}(z) & = \int\limits_0^z dz' \, \frac{Y(z')}{y(z')} , \quad
U_{+}(z) = \int\limits_z^1 dz' \, \frac{Y(1) - Y(z')}{y(z')} \,.
\end{align}
The subscript $\sigma_x$ in Eq (\ref{eq:Tx_main}) depends on $x$ as
being determined by the sign of difference $x - x_T$: $\sigma_x =+$ for
$x > x_T$ and $\sigma_x =-$ for $x < x_T$.
The functions $Y(z)$ and $U_\pm(z)$ in Eq. (\ref{eq:Tx_main}) can be
easily computed for a given profile $h(x)$ (or $y(z)$) either
analytically (see Table \ref{tab:Domains}) or numerically.  For
instance, in the simplest case of the rectangular domain, $h_{\pm}(x)
= \pm h_0/2$, we get
\begin{equation}
\label{M:e21}
T(x) = \left\{ \begin{array}{l l} \displaystyle \frac{x_T^2 - x^2}{2D} + \frac{l}{\kappa} & ~~(0 \leq x \leq x_T), \\
\displaystyle  \frac{(x - x_T)(2l - x_T - x)}{2D} + \frac{l}{\kappa} & ~~(x_T \leq x \leq l), \\ \end{array} \right. 
\end{equation}
Other examples are summarised in Table \ref{tab:Domains} and presented
below.

The explicit Eq. (\ref{eq:Tx_main}) constitutes the main result of the
paper and includes two terms.  The first (diffusion) term is
independent of the size of the target and is related to the time
required for a Brownian particle to arrive at the proximity to the
target from its starting position.  For this reason, the contribution
of this term is small when $x \approx x_T$, i.e. when the particle
initial position is near the target.  The second (reaction) term in
Eq.~(\ref{M:e21}) describes the particle absorption by the target when
the particle starts in its vicinity.  This term diverges
logarithmically as the target size decreases and thus dominates in the
small target limit.

In many applications, the starting point is not fixed but uniformly
distributed inside the domain.  In this setting, one often resorts to
the surface-averaged MFPT:
\begin{equation}
\overline{T} = \frac{1}{S} \int\limits_{\Omega} dx dy \, T(x,y).
\end{equation}
Substituting our approximate solution (\ref{eq:Tx_main}), we get an
explicit approximation for $\overline{T}$:
\begin{equation}
\overline{T}_{\rm app} = \frac{l^2}{D} \int\limits_0^1 dz \, \frac{y(z)}{Y(1)}
\bigl(U_{\sigma_{zl}}(z_T) - U_{\sigma_{zl}}(zl)\bigr) + \frac{l}{\kappa} \frac{Y(1)}{y(z_T)} \,,
\end{equation}
where $z_T = x_T/l$.  The last integral can be evaluated by using
Eq. (\ref{eq:U}).  After elementary but lengthy computations, we get
\begin{equation}  \label{eq:Tav}
\overline{T}_{\rm app} = \frac{l^2}{D} \biggl(U_{-}(z_T) + U_+(z_T) - C_y \biggr) + \frac{l}{\kappa} \frac{Y(1)}{y(z_T)} \,,
\end{equation}
where
\begin{equation}  \label{eq:C}
C_y = \int\limits_0^1 dz \, \frac{Y(z)(Y(1)-Y(z))}{Y(1) y(z)} 
\end{equation}
is the shape-dependent constant.

\section{Conditions of validity}

The main conditions that limit the range of validity of the proposed
approximation come from the two underlying assumptions: (i) a relative
smallness of the target with respect to all dimensions of the system
($s/h \ll 1$), and (ii) introduction of the effective trapping rate
(along the domain) that can adequately characterise the target.  The
effective trapping rate will be formed at some distance from the
target (since near the target it varies at much smaller scale $\sim s
\ll h$) and this imposes some restriction on the elongation of the
domain, as well as on the relative position of the starting point of
the particle (e.g., the approximation may be inaccurate if the
starting position is in the same cross-section as the target).  In
line with the previous studies \cite{Dagdug_2015}, this approximation
is expected to provide reasonable estimations of the MFPT even when
the target is not infinitesimally small and the longitudinal
separation between the starting point of the particle and the target
is of the order of the domain height.

A minor (geometrical) constraint is related to the proximity of the
target to the reflecting boundary.  If the distance between the target
and the boundary is smaller than the half-length $s/2$ of the
equivalent absorbing interval, the result will be indistinguishable
from the scenario when the target is touching the boundary.  Since we
assume that the target is relatively small, $s/h \ll 1$, this
limitation is insignificant.

More quantitative criteria for the validity of the proposed framework
will be established below by numerical simulations.

\section{Numerical Simulations}

\begin{figure}
\centering
\includegraphics[width=88mm]{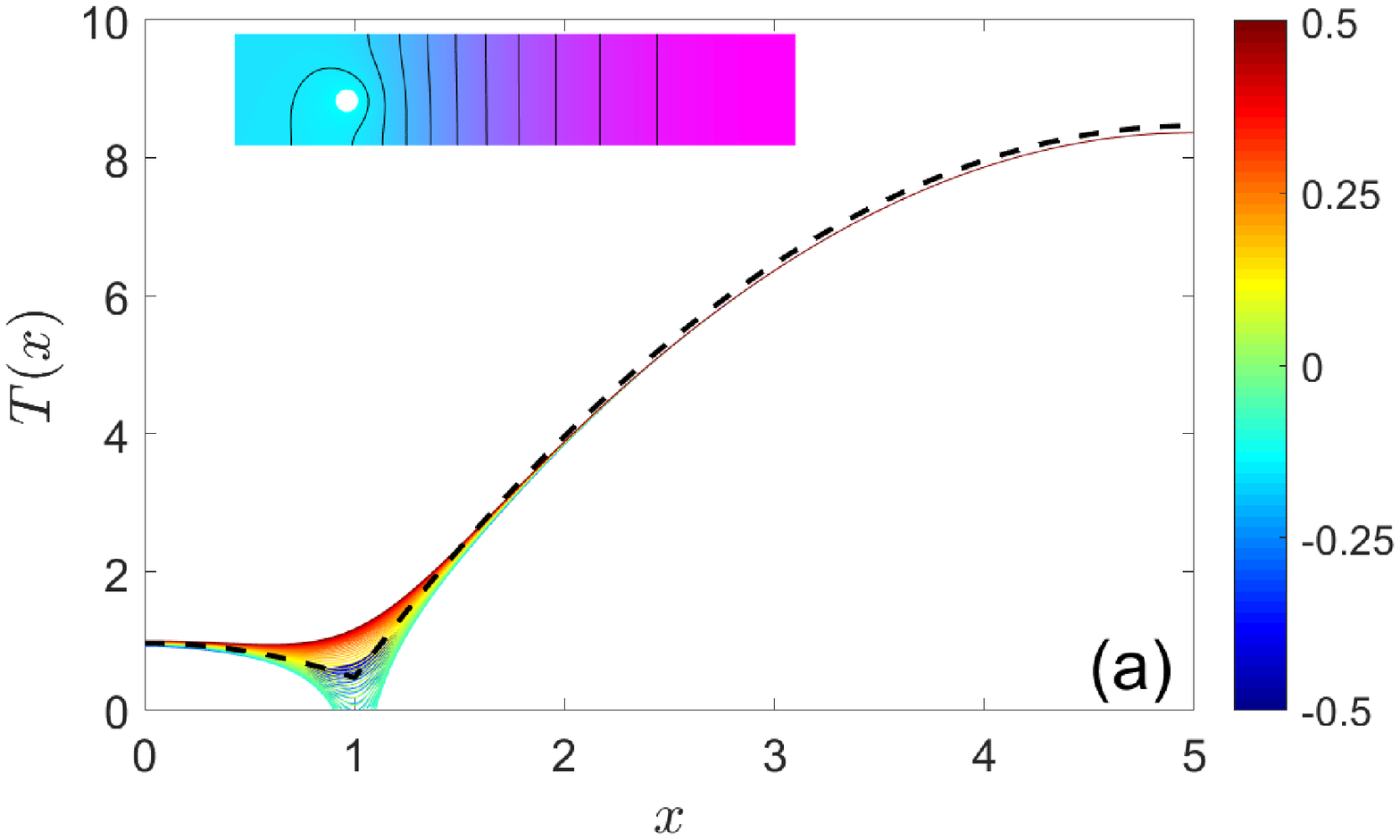}
\includegraphics[width=88mm]{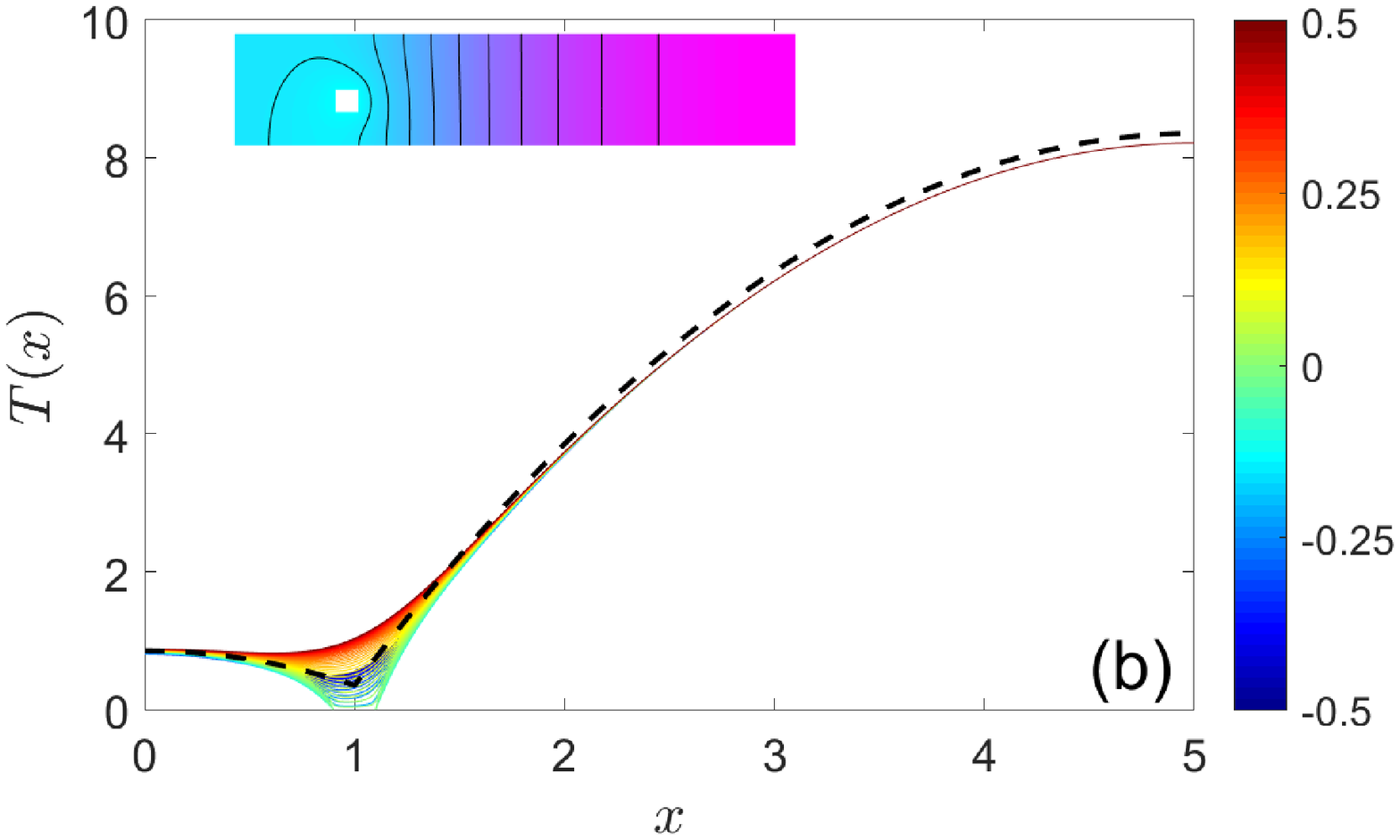}
\caption{
MFPT to a circular target of radius $\rho = 0.1$ {\bf (a)} and to a
square target with edge length $a = 0.2$ {\bf (b)}, which is located
at $(1,-0.1)$ inside an elongated rectangle $(0,5)\times
(-\tfrac{1}{2},\tfrac{1}{2})$ with reflecting boundary, with $D = 1$.
Comparison between our approximation Eq.~(\ref{M:e21}) (shown by black
dashed line) and the FEM solution (coloured lines) as a function of $x$
for 64 equally spaced $y$, from $y = -\tfrac{1}{2}$ (dark blue) to $y
= \tfrac{1}{2}$ (dark red).  Inset show the FEM solution $T(x,y)$ as
coloured contour plots. }
\label{fig:rectangle}
\end{figure}

We validate our analytical approximation for the MFPT,
Eq. (\ref{eq:Tx_main}), by comparison with the results of a direct
numerical solution of the boundary value problem (\ref{I:e1}) by means
of a Finite Element Method (FEM) solver implemented in Matlab PDEtool.
First, we calculated MFPT in the rectangular domain for a circular
target of radius $\rho$, so the length of the equivalent absorbing
interval is $s = 4\rho$.  Figure \ref{fig:rectangle}(a) shows the
MFPT, $T(x,y)$, for a circular target of radius $\rho = 0.1$ located
at $(1,-0.1)$ inside an elongated rectangle $\Omega = (0,5)\times
(-\tfrac{1}{2},\tfrac{1}{2})$ with reflecting boundary (i.e., $l = 5$
and $h_0 = 1$).  In line with the above comments, our analytical
approximation provides good estimates of the MFPT, except for the
cases when the initial position of the particle is very close to the
target (less than $h_0$).  This condition provides a quantitative
criterion for applicability of this analytical framework.

Next we analyse the effect of the target shape on the MFPT.  To this
end we consider the MFPT to a small absorbing square of side $a$ in
the same rectangular domain, for which the length of the equivalent
absorbing interval is $s = a \, \Gamma^{2}(1/4)/{\pi}^{3/2} \approx
2.36\, a$.  Figure \ref{fig:rectangle}(b) shows a good agreement
between our approximation and the FEM solution for a square of side $a
= 0.2$ at the same location inside the same rectangle as in Fig.
\ref{fig:rectangle}(a).  One can see that the target shape is
correctly captured via its conformal radius.

\begin{figure}
\centering
\includegraphics[width=88mm]{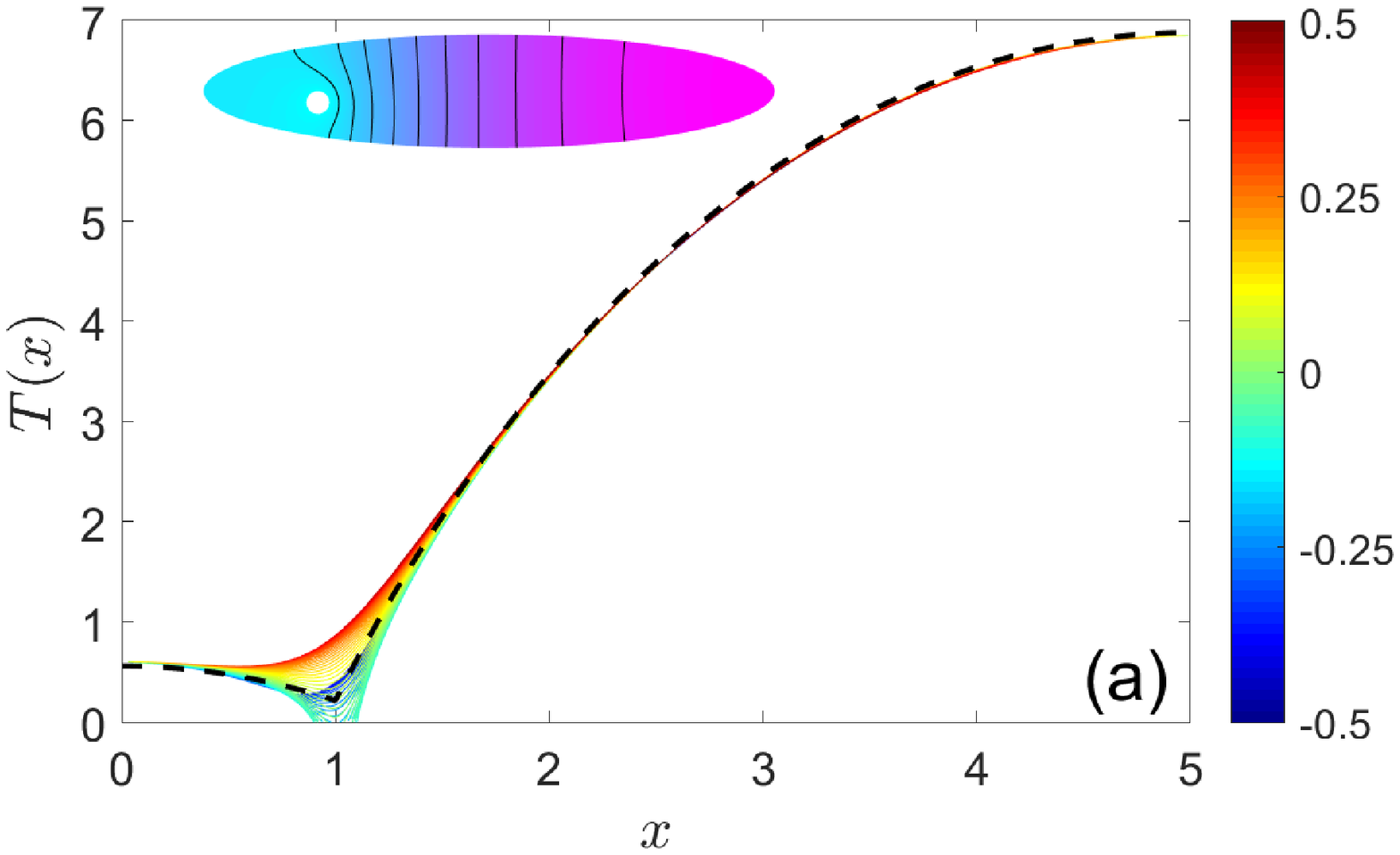}
\includegraphics[width=88mm]{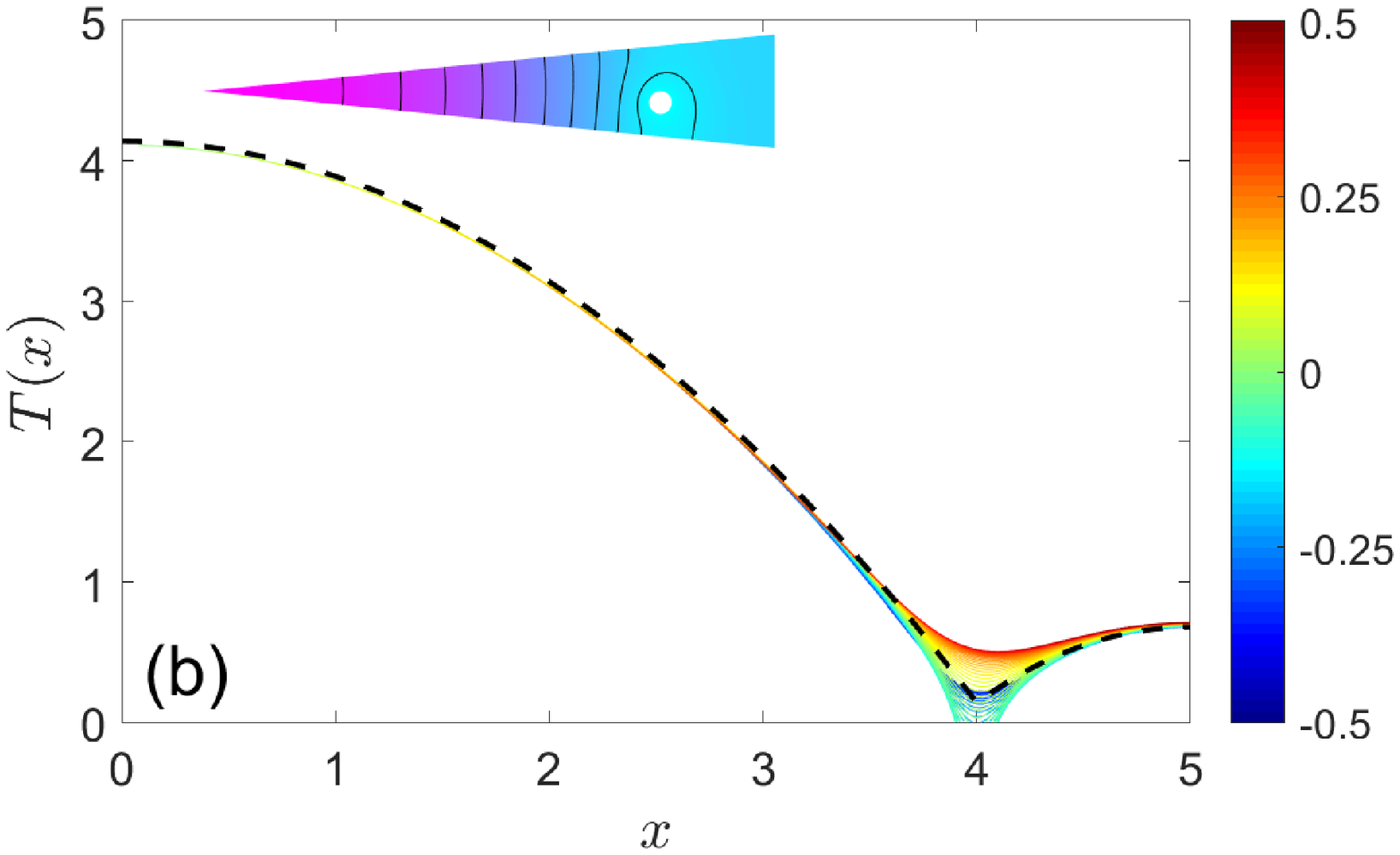}
\includegraphics[width=88mm]{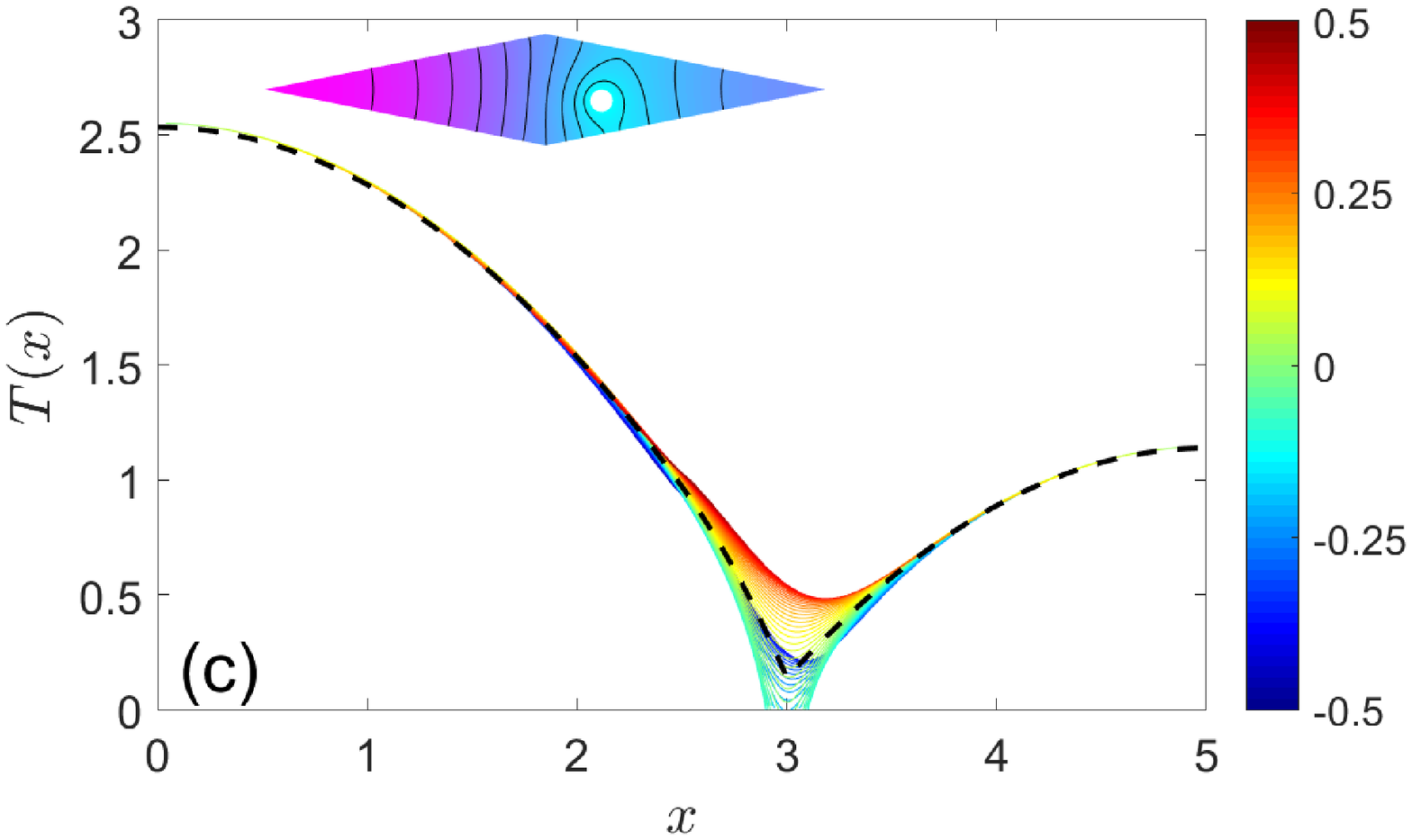}
\caption{
MFPT $T(x,y)$ to a circular target of radius $\rho = 0.1$ inside an
elongated domain with reflecting boundary, inserted into a rectangle
$(0,5)\times (-\tfrac{1}{2},\tfrac{1}{2})$, with $D = 1$ (obtained by
FEM).  {\bf (a)} Ellipse, $(x_T,y_T) = (1,-0.1)$; {\bf (b)} Triangle,
$(x_T,y_T) = (4,-0.1)$; {\bf (c)} Rhombus, $(x_T,y_T) = (3,-0.1)$.  64
coloured curves represent $T(x,y)$ as a function of $x$ for 64 equally
spaced $y$, from $y = -\tfrac{1}{2}$ (dark blue) to $y = \tfrac{1}{2}$
(dark red).  Black dashed line shows the approximate solution
(\ref{eq:Tx_main}), with explicitly found functions $Y(z)$ and
$U_{\pm}(z)$ in Table \ref{tab:Domains}.  Inset shows $T(x,y)$ inside
each domain.}
\label{fig:ellipse}
\end{figure}

To validate the proposed framework for other elongated domains we
calculated the MFPT in three domains of different shape (an ellipse, a
triangle, and a rhombus), for which we kept the same aspect ratio as
before: $h_0/l = 0.2$.  A circular target of radius $\rho = 0.1$ is
located in different positions inside these domains.  Figure
\ref{fig:ellipse} illustrates an excellent agreement between our
approximation in Eq. (\ref{eq:Tx_main}) and numerical solutions for
all these domains.  We also checked that the relative error of the
explicit approximation (\ref{eq:Tav}) for the surface-averaged MFPT,
$\overline{T}$, does not exceed $2\%$ for all these examples (see
Table \ref{tab:Tav}).

\begin{table}[t!]
\begin{center}
\begin{tabular}{|c|c|c|c|}  \hline
Domain & $\overline{T}$ & $\overline{T}_{\rm app}$ & relative error\\  \hline
Rectangular (circle) & 4.7481  & 4.7961 & $1\%$ \\
Rectangular (square) & 4.6128  & 4.6855 & $2\%$ \\ 
Triangular  & 1.5567  & 1.5522 & $0.3\%$ \\
Rhombic     & 1.2096  & 1.1888 & $2\%$ \\
Elliptic    & 4.0403  & 4.0340 & $0.2\%$ \\  \hline
\end{tabular}
\caption{
The surface-averaged MFPT, $\overline{T}$, estimated from the
numerical solution of the original problem (solved by a FEM), and its
explicit approximation $\overline{T}_{\rm app}$ from
Eq. (\ref{eq:Tav}), for five examples shown in
Figs. \ref{fig:rectangle} and \ref{fig:ellipse}.}
\label{tab:Tav}
\end{center}
\end{table}

\section{Conclusion}

We obtained a simple formula (\ref{eq:Tx_main}) for the MFPT to a
small absorbing target of an arbitrary shape in an elongated planar
domain with slowly changing boundary profile.  This formula expresses
MFPT in terms of dimensions of the domain, the form and size of the
absorbing target and its relative position inside the domain.  We
validated our analytical predictions by numerical simulations and
found excellent agreement.  Indeed, if the initial position of the
particle and the target location are well separated (more than half of
the height of the domain at the target location $x_T$) then the
numerical and analytical results are almost indistinguishable; but
even for closer separations the analytical predictions are still
reasonable, see Figs. \ref{fig:rectangle} and \ref{fig:ellipse}.  The
proposed expression for the MFPT is a useful tool for some rapid
practical estimations as well as for validation of complex numerical
models of particle diffusion in geometrically constrained settings.

Future work may involve an extension of the proposed framework to more
complex geometries (an elongated domain with an arbitrary piecewise
boundary) or an extension to the three-dimensional settings.  The main
challenge in the three-dimensional case consists in finding an
appropriate expression for the trapping rate of the target, i.e., a 3D
generalisation of Eq.~(\ref{M:e9}).

\begin{acknowledgments}
D.~S.~G. acknowledges a partial financial support from the Alexander
von Humboldt Foundation through a Bessel Research Award.
A.~T.~S. thanks Alexander M. Berezhkovskii and Paul A. Martin for many
helpful discussions.
\end{acknowledgments}

\end{document}